\documentclass[twocolumn,showpacs,preprintnumbers,amsmath,amssymb,prb,aps]{revtex4-1} 
\usepackage{epsfig}
\usepackage{epstopdf}
\usepackage{multirow}
\usepackage{graphicx}
\usepackage{dcolumn}
\usepackage{bm}
\usepackage{textcomp}
\usepackage{array}
\usepackage{romannum}
\usepackage{csquotes}
\usepackage{subcaption}
\usepackage{booktabs}
\usepackage{amsmath}
\usepackage{mathtools}
\usepackage{enumitem}
\usepackage{hyperref}
\hypersetup{
    colorlinks=true,
    linkcolor=blue,
    filecolor=magenta,      
    urlcolor=blue,
    pdftitle={Abhishek Pandey},
    }
    
\begin{document}
\title{Exploring the effect of strong electronic correlations in Seebeck Coefficient of the NdCoO$_3$ compound : Using experimental and DFT+$U$ approach}
\author{Abhishek Pandey$^{1}$}
\altaffiliation{ \url{ vivekiit.19.mandi@gmail.com}}
\author{Sudhir K. Pandey$^{1}$}
\altaffiliation{ \url{sudhir@iitmandi.ac.in}}
\affiliation{$^{1}$School of Mechanical and Materials Engineering, Indian Institute of Technology Mandi, Kamand - 175075, India}

\date{\today}

\begin{abstract}
The presence of complexity in the electronic structure of strongly correlated electron system NdCoO$_3$ (NCO) have sparked interest in the investigation of its physical properties. Here, we study the the Seebeck coefficient ($\alpha $) of NCO by using the combined experimental and DFT+$U$ based methods. The experimentally measured $\alpha $ is found to be $\sim$ 444 $\mu V/K$ at 300 K, which decreases to 109.8 $\mu V/K$ at 600 K. In order to understand the measured $\alpha $, we have calculated the PDOS and band structure of the NCO. Furthermore, the calculated occupancy of 6.4 for Co $3d$ orbitals and presence of  large unoccupied O $2p$ states indicate the covalent nature of the bonding. Apart from this, the maximum effective mass is found to be 36.75 (28.13) for the spin-up (dn) channel in conduction band indicates the n-type behaviour of the compound in contrast to our experimentally observed p-type behaviour. While, the calculated $\alpha $ at the temperature-dependent chemical potential ($\mu$) at 300 K shows the p-type behaviour of the compound. Fairly good agreement is seen between the calculated and measured values of $\alpha $ at U$_f$$_f$ = 5.5 eV and U$_d$$_d$ = 2.7 eV. The maximum power factor (PF) is found to be 47.6 (114.4) $\times$  $10^1$$^4 $ $\mu $$W$$K^-$$^2$$cm^-$$^1$$s^-$$^1$ at 1100 K, which corresponds to p (n)-type doping of $\sim$ 1.4 (0.7) $\times$ 10$^2$$^1$ cm$^-$$^3$. This study suggests the importance of strong on-site electron correlation in understanding the thermoelectric property of the compound.
 
\end{abstract}

\maketitle

\section{Introduction}

Thermoelectricity is an environmental friendly physical phenomena that allows for the invertible conversion of thermal and electrical energy.  
By utilising waste heat, it opens up a green alternative for energy generation \cite{ZhangXiao,Yang,Diez}. 
The performance of thermoelectric material is defined by the dimensionless parameter figure of merit ($ZT$)  \cite{Yanzhongzt,LaLonde}.
\begin{equation}
  ZT=\frac{\alpha ^2\sigma T}{\kappa _e+\kappa _l}
\end{equation}
where, $\alpha $ is Seebeck coefficient, $\sigma $ is electrical conductivity, $T$ is absolute temperature, $\kappa $$_l$ is lattice thermal conductivity  and $\kappa $$_e$ is electronic thermal conductivity. 
In the formula of $ZT$, $\alpha ^2\sigma $ is known as power factor, which plays significant role in deciding the potential of TE materials \cite{Nolasbook}. 
To achieve a high $ZT$, the material should have high power factor which is directly proportional to the square of $\alpha $.
Therefore, the study of the $\alpha $ becomes useful for any application purpose.

In recent years, fascinating structural transitions, transport properties, and large $\alpha $ of LnCoO$_3$ \cite{HASHIMOTO2009246,F29757101885,MOON200170,Takami_2004,He,Ivanova,Yamaguchi,topsakal2016first,DUDNIKOV202017987, Orlov, Saha,Ali,KIANIPOUR2022100506} series have attracted the scientists.
Furthermore, these compounds have gained attention due to high temperature oxidation resistance and simple synthesis technique.
From the rare earth cobaltates family, NdCoO$_3$ (NCO) is also gaining traction due to high $\alpha $ and power factor\cite{HASHIMOTO2009246,DUDNIKOV202017987,Tealdi,Ashutosh}.
Dudnikov $et$ $al$ \cite{DUDNIKOV202017987} and Ashutosh \cite{Ashutosh} have reported the similar nonmonotonic behaviour of $\alpha $ for the compound. 
The measured value of $\alpha $ found to be $\sim$ -180, 379, and 108 $\mu $V/K at 300, 450, and 600 K, respectively in Dudnikov $et$ $al$ work. 
Additionally, Tealdi $et$ $al$ \cite{Tealdi} reported the $\alpha $ from 300-1100 K of the NCO. From Tealdi $et$ $al$ work, measured value of $\alpha $ was found to be $\sim$ 284, 153, and 82 $\mu $V/K at 300, 450, and 600 K, respectively.
In their work, solid state reaction technique was used for the synthesis of the compound.
However, instead of having similar synthesis process and formation of compound in same space group, Dudnikov $et$ $al$ and Ashutosh have reported both n- and p-type behaviour, whereas Tealdi $et$ $al$ reports only p-type behaviour of the compound. That is, the nature and magnitude of the reported $\alpha $ in their work found to be different.
 
 The NCO have two types of valence-electron orbitals. The first is extended s and p orbitals that overlap in order to form the broad free-electron type bands which are treated well by DFT\cite{Kohn,Hohenberg} approach.
In the second type, $3d$ and $4f$ orbitals are well localized, and in solids, they form
 narrow bands.
Moreover, these are pinned to Fermi energy and weakly hybridize with s and p orbitals\cite{Steiner}. Also it is well known that these orbitals are not treated well through DFT. 
In the computational study of NCO, Mehmet $et$ $al$ \cite{topsakal2016first} have reported the dispersion curve for the compound using LDA+$U$ and GGA+$U$ calculations.
Since, it is known that NCO is a strongly correlated electron system (SCES \cite{lundin2005theory}), with partially filled and highly localised Nd $4f$ and Co $3d$ orbitals.
Also, the electronic states in the vicinity of $E_f$ plays very significant role in transport property of the materials. 
 Therefore, it is expected to have substantial role for Nd $4f$ and Co $3d$ orbital in the compound's electronic and transport properties.  
Furthermore, to the best of our knowledge, no such theoretical investigation beyond DFT have been done to explore the reasons for large $\alpha $ of the compound.
Analysing the experimental data of Ashutosh, Dudnikov $et$ $al$, Tealdi $et$ $al$ and theoretical study of  Mehmet $et$ $al$,
it can be concluded that a systematic investigation and understanding of $\alpha $ for the compound is lacking in the literature.
The computational apporach is helpful in the search of new materials and analysing the experimental results. 
Density functional theory (DFT) is a prominent computational approach for predicting the ground state electronic structure of different materials.
However, DFT cannot describe the electronic structure of the  compounds having highly localized and partially filled d or f-orbitals.
In such scenario, DFT+$U$ can be used, in which on-site coulomb interaction U holds the responsibility for overcoming the problem.
Therefore, it becomes important to study the $\alpha $ of the NCO using experimetnal and beyond DFT techniques. 
Thus, in this work we have explored the extent to which experimentally observed $\alpha $ of NCO can be explained by DFT+$U$.

Considering the above mentioned aspects,
we have investigated the $\alpha $ using experimental and theoretical approaches for the NCO compound. 
 The experimentally measured value of $\alpha $ varies from $\sim$ 444 to 109.8 $\mu $V/K in the temperature range of 300-600 K.
The positive sign of $\alpha $ in the whole temperature range indicates the p-type behaviour of the compound.
In order to analyse the measured $\alpha $ values, we have performed the electronic structure calculation in DFT+$U$ framework. 
The calculated occupancy of Co $3d$ states and unoccupied O $2p$ states indicate the presence of covalent bonding in the NCO.
Effective mass ($m^*$) of the electrons and holes have been estimated from electronic dispersion around $\Gamma $-point along the $\Gamma $-other high symmetric directions. 
The large effective mass ($m^*$) around $\Gamma $ point along $\Gamma $-$Z$ direction suggest the sign of $\alpha $ to be negative for the compound. However, the calculated value of $\alpha $ with considering the temperature-dependence in chemical potential ($\mu$) shows the p-type behaviour of the compound, which is same as our experimental result.  
We used the semi-classical Boltzmann theory for the transport calculations.
 Fairly good agreement is observed between the calculated and experimental data for $\alpha $ value at U$_f$$_f$ = 5.5 eV and U$_d$$_d$ = 2.7 eV.  
The maximum power factor is calculated for both n and p-type doping.
These findings suggests the importance of strong electronic correlation in the $\alpha $ and consequently the power factor of the compound.
\section{Experimental and Computational Details}
\begin{figure} 
\begin{center}
\includegraphics[width=0.45\textwidth,height=6cm]{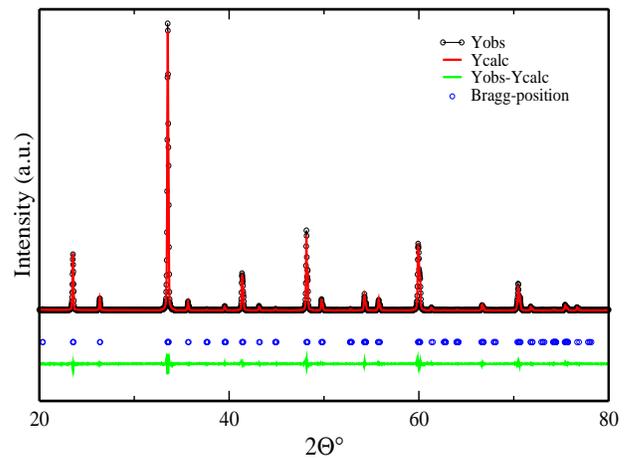} 
\caption{$\label{Figure1}$\small{Room temperature X-ray diffraction of NdCoO$_3$.
}}
\end{center}
\end{figure}

NdCoO$_3$ sample was synthesized by using the combustion technique \cite{Mukasyan}. The initial ingredients Nd(NO$_3$).6H$_2$O and Co(NO$_3$).6H$_2$O were added in DI-water. 
This mixture was stirred about 1/2 hour with the help of magnetic stirrer. Then after glycine was added slowly to the solution and again stirred for 2 hours.
Resultant solution was heated at 100$^{\circ}$C for some time, then raised the temperature to 150$^{\circ}$C and waited for the water to evaporate to form the gel. Further heating leads the gel to catch fire, which results in a powdered sample. This primary powder was calcinated at 900°C for 24 h.
 Obtained powder was pressed to pellets and kept in furnace at 1000 $^{\circ}$C for 24 hours to harden the pellet.
 In order to determine the crystal structure of the sample, X-ray diffraction(XRD) technique was used at the room temperature. 
For producing the diffractogram, Cu $\kappa$$_\alpha$ radiation was used. 
Resultant diffractogram was recorded with step counting mode 0.02 in the range of $20^\circ-80^\circ$. Lattice and structural parameters were analyzed using the Rietveld refinement method \cite{Rietveld}. 
Temperature dependent $\alpha $ measurement carried out using the home-made instrument\cite{sk2022instrument} setup in the temperature range of 300-600 K .

The ground state spin unpolarized (SUP) and spin polarized (SP) DFT and DFT+$U$ calculations have been done using the augmented plane wave plus local orbitals (APW+lo) method as implemented in the WIEN2K code \cite{tran2018wien2k}. 
The PBEsol \cite{Perdewpbesol} functional is used as the exchange and correlation (XC) part within generalized gradient approximation (GGA). 
The muffin-tin radii of sphere for Nd, Co, O atoms are set to 2.17, 1.83, 1.63 Bohr, respectively. 
Self-consistent energy convergence criteria was fixed to 0.1 mRy/cell.  
Temperature dependent electronic transport coefficient are calculated using the BoltzTraP\cite{boltz} package based on Boltzmann transport theory \cite{Allen1996BoltzmannTA}.
The electronic transport coefficient calculations have been carried out on the k-mesh of  $50\times 50\times 50$.

\begin{figure}
  
\begin{center}
\includegraphics[width=0.35\textwidth,height=7.0cm]{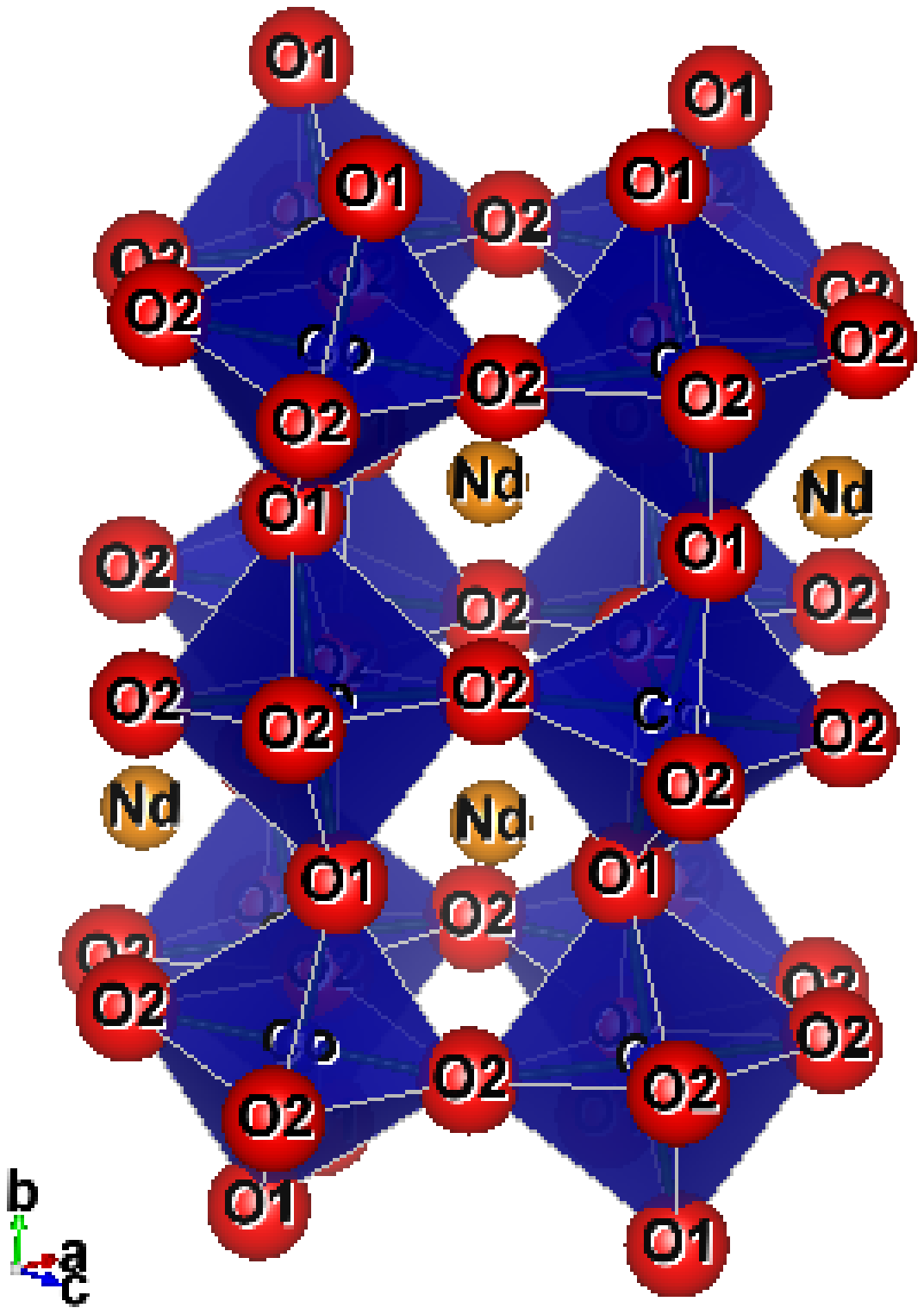}
\caption{\small{Distorted orthorhombic crystal structure of NdCoO$_3$.
}}
\end{center}
\end{figure}

\begin{table}
\begin{center}
\caption{\label{tab:table1}%
\small{Bond length and angle for orthorhombic structure of NdCoO$_3$}}
\setlength{\tabcolsep}{6pt}
\begin{tabular}{lcc}
\colrule 
\multicolumn{1}{c|}{Lattice Parameter} & \multicolumn{1}{c}{Wyckoff Positions} \\
\colrule
\midrule
\multicolumn{1}{c|}{a = 5.33707(15)}&\multicolumn{0}{c}{Nd[0.96494(17), 0.25(00), 0.00611(35)]} \\
\multicolumn{1}{c|}{b = 7.55316(24)} & \multicolumn{0}{c}{Co[0.0, 0.0,  0.5]} \\
\multicolumn{1}{c|}{c = 5.34779(19)} & \multicolumn{0}{c}{O1[0.49356(20), 0.75(00), 0.09240(00)]} \\
\multicolumn{1}{c|}{}& \multicolumn{0}{c}{O2[0.29174(23),0.54910(16),0.70270(27)]} \\

\colrule

Co-O1  &  1.95438(4) \\
Co-O2  &  1.874(14)\\
Co-O2  &  1.968(14) \\

\colrule
&\textrm{Bond angle}\\
\colrule
Co-O1-Co &  150.60(5)\\
Co-O2-Co &  158.9(8) \\
\colrule
\end{tabular}
\end{center}
\end{table}

\section{Results and Discussion}
Figure 1 represents the X-ray diffraction pattern at room temperature with fitted patterns of NCO compound. 
The analysis of the XRD pattern have been done by using the Rietveld refinement method by means of fullprof software package.
 Data for the structure of NCO are taken from available data in the literature for fitting the pattern \cite{ateia2019synthesis,Orlov}. 
  In the figure open circles (black) are showing the experimental data while solid (red) curve is showing the calculated pattern.
   Moreover, the blue squares are showing the Bragg positions and the curve at the bottom is representing the difference between experimental and calculated pattern.   
No impurity peaks are found for the compound.
 The ratio of the weighted profile factor (R$_w$$_p$) to the expected weighted profile factor ($R_e$$_x$$_p$) indicates the goodness of fitting for structure. It has a value of 1.7 for our compound. 
The good match of the reported ones with the refined lattice parameters, and absence of any impurity peak in the XRD pattern confirms the single phase of the sythesized compound.
Compound shows the Orthorhombic structure with Pnma (No. 62) space group.
 The obtained refined Wyckoff positions of Nd, Co, O1, and O2 are found to be 4c[0.96494(17), 0.25, 0.00611(35)], 4b[0.0, 0.0, 0.5],  4c[0.49356(20), 0.75, 0.09240(00)], and O2 - 8d [0.29174(23), 0.54910(16), 0.70270(27)], respectively. 
The lattice parameter obtained from the refinement is $a$ = 5.33707(15), $b$ = 7.55316(24), $c$ = 5.34779(19) \AA. 
Refined lattice parameters and atomic positions are used to estimate the Co-O (Co-O-Co) bond length (bond angle).  Structure of the compound is shown in figure 2. 

Figure shows that O2 is in the apical position and O1 is in the equatorial position of the octahedral geometry formed around the Co. 
The measured bond  angle between Co-O1 [O2]-Co is found to be 150.60(5) [158.9(8)].
Co-O2-Co angle holds the responsibility of CoO$_6$ octahedra rotation about the ac plane. While, Co-O1-Co bond angle provides the measure of deviation from the idealized 180$^\circ$ along b-axis.
The  bond distance for Co-O1 (O2) is found to be 1.954 (1.874 and 1.968) \AA. 
The three Co-O bond length can be distinguished from figure as Co-O1 at the apical position and two bond length of Co-O2 at the equatorial plane of the octahedra. 
Therefore, depending on the differences between these bonds, the environment of cobalt can be attributed to distorted octahedra.
This distortion in the crystal structure can be attributed to the strong link to the electronic properties of the compound. 
Therefore, proper analysis of the the electronic partial density of states (PDOS) and dispersion of the compound is required to explain the electronic transport properties.


\begin{figure}
\includegraphics[width=0.8\linewidth, height=5cm]{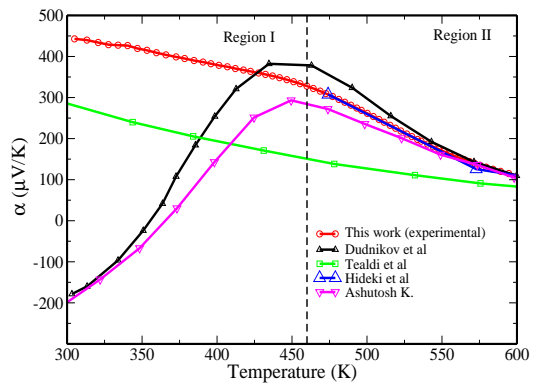} 
\caption{\small{Temperature dependence of measured Seebeck coefficient $\alpha $.}}
\label{fig:image2}
\end{figure}
Figure 3 shows the experimentally measured $\alpha $ of this work in the temperature range of 300 to 600 K alongwith the work of Ashutosh\cite{Ashutosh}, Hideki $et$ $al$\cite{HASHIMOTO2009246}, Dudnikov $et$ $al$.\cite{DUDNIKOV202017987} and Tealdi $et$ $al.$ \cite{Tealdi}. 
Negative and positive value of $\alpha $ comes from electrons and holes dominating contribution in the material.
In our work, the positive $\alpha $ in temperature range of 300-600 K indicates the p-type behaviour of the compound.  
The measured value of $\alpha $ at $300$ (600) K is found to be $\sim$444 (109.8) $\mu V/K$.  
With increase in temperature from 300-600 K, the $\alpha $ value continuously decreases. The rate of decrement of the $\alpha $ below 460 K found to be $\sim$ -0.73 $\mu V/K^2$, which increases to $\sim$ -1.56 $\mu V/K^2$ in the temperature range of 460-600 K. 
To compare our measured $\alpha $ to those of Dudnikov $et$ $al.$\cite{DUDNIKOV202017987}, Ashutosh \cite{Ashutosh} and Tealdi $et$ $al.$ \cite{Tealdi}, the temperature range can be divided into two regions accounting the rate change of $\alpha $ around 460 K [ where, region-\Romannum{1}(\Romannum{2}) is from : 300-460 (460-600) K].
In region-\Romannum{1}, Tealdi $et$ $al.$ \cite{Tealdi} have reported the p-type behaviour of the compound, which is same to our finding.
The difference between the $\alpha $ in our and Tealdi $et$ $al.$ work is found to be $\sim$ 155 $\mu V/K$ at room temperature and remains almost constant up to $\sim$ 460 K. 
In region \Romannum{2}, the difference in $\alpha $ decreases with increase in temperature to $\sim$ 20 $\mu V/K$ at 600 K.
Furthermore, figure shows that Dudnikov $et$ $al.$ and Ashutosh obtained the negative value of $\alpha $ about room temperature in region \Romannum{1}. The negative $\alpha $ represents the n-type behaviour in the compound, which is in contradiction of our and Tealdi $et$ $al.$ work. 
With increase in temperature, $\alpha $ in the Dudnikov $et$ $al.$ and Ashutosh work increases and becomes positive about $\sim$ 355 K. The change in sign of slope in $\alpha $ v/s T is observed about $\sim$ 460 K, which leads to decrease in $\alpha $ on further increase in temperature. 
In addition, Dudnikov $et$ $al.$\cite{DUDNIKOV202017987}, Ashutosh \cite{Ashutosh} and Tealdi $et$ $al.$ \cite{Tealdi} used the similar solid state reaction method to prepare the compound. Therefore, the observed difference in the value of $\alpha $ in region-\Romannum{1} might be attributed to different synthesis conditions.
In addition, we can say that $\alpha $ is highly sensitive in the temperature range of 300-450 K.
Furthermore, our finding indicate that synthesis of NCO by combustion method is most suited for attaining high $\alpha $ in region-\Romannum{1}.
Each work has reported roughly linear decrease with similar $\alpha $ at high temperature of 460 K and above in region-\Romannum{2}. 
This linear decrease can be attributed to Insulator-Metal Transition \cite{Saha,Itoh,DUDNIKOV202017987,Kn} in the compound. 

\begin{figure*}
\begin{center}
\includegraphics[width=0.90\textwidth,height=8.50cm]{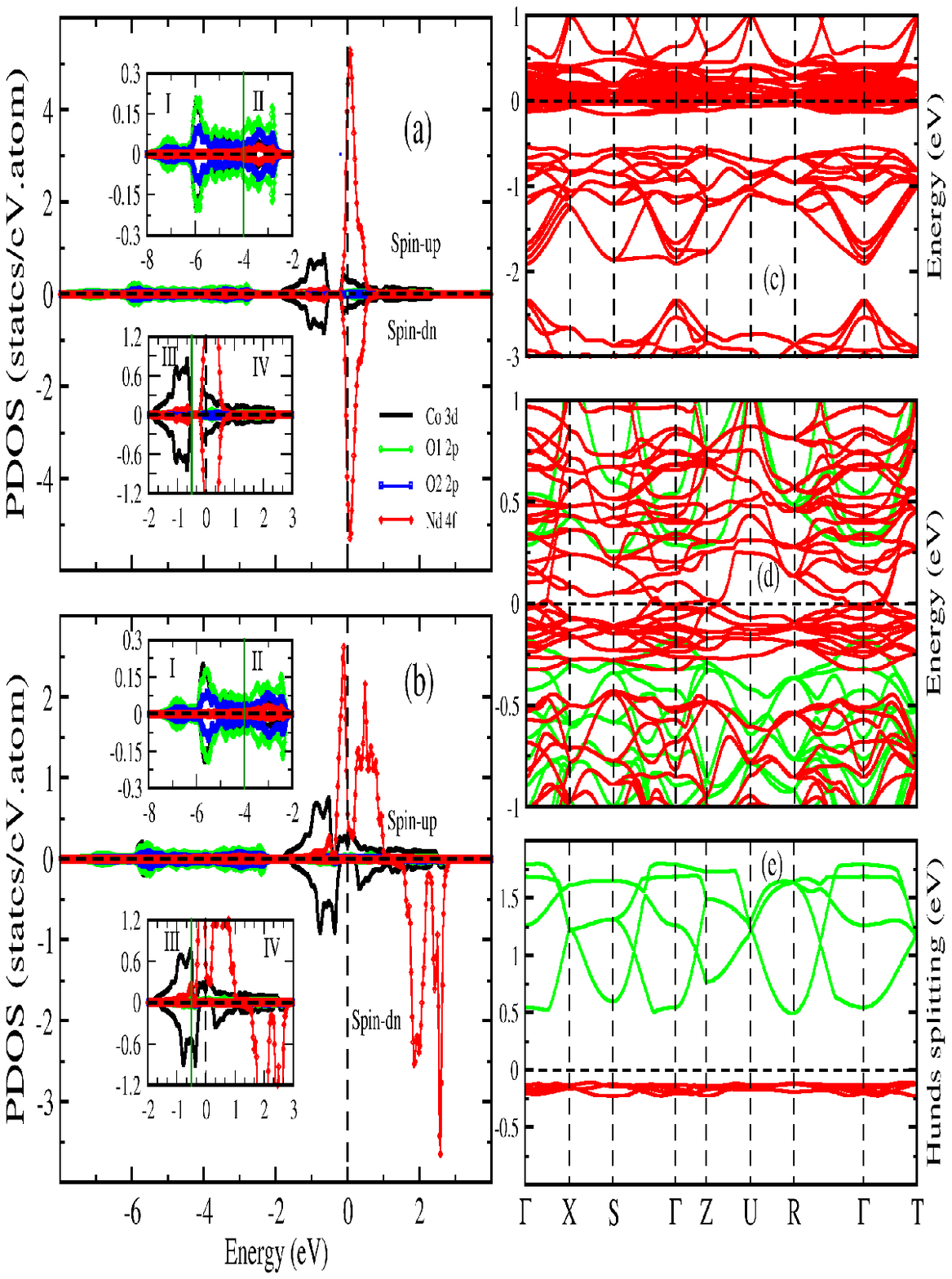} 
\caption{$\label{Figure3}$ \small{Partial density of states (PDOS) of (a) spin-unpolarized (SUP), (b) spin-polarized (SP) using Density Functional Theory (DFT). Dispersion curve of (c) SUP, (d) SP (Spin-up red and Spin-dn green) using DFT, (e) Showing the Hunds splitting.
}}
\end{center}
\end{figure*}
For the understanding of experimentally measured $\alpha $, electronic structure calculations have been carried out.
Figure 4a(4b) represents the spin-unpolarized SUP (spin-polarized SP) partial density of states (PDOS) of the NCO by using the density functional theory \cite{Kohn,Hohenberg} (DFT). 
For the detailed analysis of PDOS, the energy window has been divided into four regions \Romannum{1}, \Romannum{2}, \Romannum{3}, and \Romannum{4} which ranges from -8 to -4 eV, -4 to -2 eV, -2 to -0.5 eV, and -0.5 to 4 eV, respectively. 
  
In the figure 4(a) region \Romannum{1}, spin-up/dn of Co $3d$, O1 (O2) $2p$, and Nd $4f$ have 0.17, 0.25 (0.12), and 0.01 states/atom, respectively. Here, Co $3d$, O1 (O2) $2p$ orbitals have almost similar contribution in the energy window.
In region \Romannum{2}, orbitals of Co $3d$, O1 (O2) $2p$, and Nd $4f$ have 0.04, 0.14 (0.08), and 0.02 states/atom, respectively. 
The features in the energy window is dominated by O1 (O2) $2p$ states with negligible contributions from other orbitals.
In region \Romannum{3}, Co $3d$, O1 (O2) $2p$, and Nd $4f$ have 0.51, 0.03 (0.01), and 0.05 states/atom, respectively. Here, one can observe that Co $3d$ with 0.51 states/atom dominated in the region with negligible contributions from other orbitals.
Nd $4f$ with 1.67 states/atom dominated the region \Romannum{4} with the finite states of Co $3d$ and negligible states of O1(O2). Most of the states of Nd $4f$ lies within narrow energy range of -0.2 to 0.8 eV with finite states of Co $3d$ (0.37 states/atom) and insignificant O1 (O2) $2p$ states have been found.

The dominating character of O$1$/O$2$ $2p$, Co $3d$, and Nd $4f$ has been found in the energy window of -4 to -2.5 eV, -1.85 to -0.5 and -0.12 to 0.51 eV, respectively, which can be attributed as their non bonding states in the region.
 The order in the energy of non-bonding states shown by each element's PDOS is in the agreement to the order of their atomic orbital energy. 
The presence of similar weight of O1(O2) $2p$, Co $3d$ orbitals in the same energy window -8 to -4 eV can be attributed as bonding states and this can be considered as a signature of strong hybridization between the Co $3d$ and O $2p$ orbitals. 
In addition, no states of any element found in the energy window of -2.34 to -1.9 eV and -0.45 to -0.25 eV indicating the presence of band gap in the energy window. 

The available experimental results indicated the insulating property of the compound. 
In contrast to experimental results, SUP solutions shows significantly large states available at $E_f$ in the figure, which demonstrates the metallic behaviour of the compound. It implies the inability of DFT in producing the correct insulating ground state.
Figure 4(b) shows spin resolved PDOS. The SP solutions in the figure of regions I and II are similar to the SUP solutions in figure 4(a). However, in these regions SP solutions shifted towards the Fermi level (E$_f$). This shifting reduces the band gap to 0.2 (from -2 to -1.8) eV for up-channel and 0.4 (from -2 to -1.6) eV for dn-channel, which is found to be 0.44 (from -2.34 to -1.9) eV for SUP solutions. 
In region \Romannum{3} of figure 4(b), the spin-up (dn) states of Co $3d$ get broadend in the energy window and extend to the region \Romannum{4}.
The band edge of Co $3d$ at -1.9 eV in SUP solutions shifted to -1.85 (-1.66) eV for spin-up (dn) states in SP solution.
The difference in band edge shift between spin-up and spin-dn states is found to be 0.21 eV and can be attributed as exchange splitting of Co $3d$ states in the region.  
 In region \Romannum{4} of figure 4(b), four significant peaks are seen for the Nd $4f$ states. From these four peaks, two peaks are from spin-up states and two peaks are from spin-dn states.
Furthermore, the spin-up (dn) states from these peaks are highly localized and dominated in the energy window of -0.38 to 1 eV (1.15 to 2.92 eV).
 It implies that Nd $4f$ spin-up (dn) states have shifted and broadened towards left (right) directions from the E$_f$ similar to Co $3d$ in region \Romannum{3}.
  The band gap of 0.2 (from -0.45 to -0.25) eV found in the energy window of SUP solutions is filled by the widening of Nd $4f$ and Co $3d$ states in the SP solutions.
Also, a band gap of 0.26 (from -0.1 to 0.16) eV is found in the vicinity of $E_f$ for dn-channel.
In contrast to the insulating behaviour of the compound revealed by experimental data, considerable presence of Nd $4f$ and Co $3d$ states are seen for the spin-up channel at the E$_f$, which indicates the compound's half-metallic behaviour.
The band edge of Nd $4f$ at -0.23 eV in SUP solutions shifted to -0.38 (1.5) eV for spin-up (dn) states in SP solutions. The difference in band edge shift for spin-up and spin-dn is 1.88 eV.

Figure 4c(4d) is showing the SUP (SP) electronic dispersion of the compound along the high symmetric directions of $\Gamma$-$X$-$S$-$\Gamma $-$Z$-$U$-$R$-$\Gamma$-$T$ in the brillouin zone (BZ).
At the $\Gamma$ point in dispersion for SUP solutions the band gap of 0.44 (from -2.34 to -1.9) eV is seen, which is consistent with the band gap seen in SUP-PDOS in the energy window. 
However, in the figure 4(c), indirect band gap of 0.39 (from -0.55 to -0.16) eV is found between $\Gamma $ and a point in $\Gamma $-$S$ direction, which is greater than the band gap of 0.2 (from -0.45 to -0.25) eV in SUP-PDOS.
In figure 4(d), the dispersion of spin-up (dn) is represented by red (green) colour.
The observed spin-up states of Nd $4f$ and Co $3d$ at E$_f$ in figure 4(b) can be seen in the dispersion curve of figure 4(d).
Previous studies have shown that the magnetic and electronic properties are very sensitive to the behaviour of dispersion at the $E_f$. 
Therefore, we choose only those bands which are crossing the $E_f$ in SUP dispersion.
In order to see the energy shift of states which are crossing the $E_f$ having band numbers 92-95 in SUP dispersion, we have taken the same 92-95 band numbers from SP dispersion of spin-up and spin-dn channels, which is shown in the figure 4 (e).
The average energy difference between spin-up and spin-dn states in these bands are approximately 1.58 eV. Moreover, at $E_f$ electrons of $4f$ are dominating and therefore this gap may be attributed to exchange splitting Hunds J value of Nd $4f$ states.
In conclusion, DFT-SP solutions give the half-metallic nature of the compound, in contrast to the experimental results of insulating behaviour.
Since, it is well known that DFT fails to properly account the electron-correlation of highly localized and partially filled $4f$ and $3d$ orbitals. 
Therefore, the non-insulating nature shown by DFT-SP may be due to presence of highly localized and partially filled Nd $4f$ and Co $3d$ orbitals in the compound.
From this, one can say that NCO is strongly correlated electron system and beyond DFT calculations are necessary to explain experimental data.

Figure 5 is representing the PDOS obtained using DFT+$U$ calculations. The U$_e$$_f$$_f$ (U-J) term for Nd $4f$ (Co $3d$) orbitals are represented as U$_f$$_f$ (U$_d$$_d$).
Singh $et$ $al$. \cite{Saurabh} have used the U$_d$$_d$ of 2.75 eV for Co $3d$ orbitals in LaCoO$_3$ compound and shown good match of calculated transport properties with the experimental results. 
Therefore, in order to analyse the strongly correlated electrons of Nd $4f$ and Co $3d$ states in NCO, we have calculated the PDOS by taking the U$_d$$_d$ of 2.7 eV (alongwith U$_f$$_f$ of 0, 2.5, 3.5, 4.5, and 5.5 eV). 

In figure 5(a), given value of U$_d$$_d$ (U$_f$$_f$) is 2.7 (0) eV.
In this case, Co $3d$ states are shifted away from $E_f$ and showing the band gap of 0.5 eV below the E$_f$. 
Also, Nd $4f$ states remained dominated around E$_f$. After including the U$_d$$_d$ value, it was expected to have no states of Co $3d$ around E$_f$ but finite states have been found. 
Since, the orbitals of Nd $4f$ and Co $3d$ are very localized and bond length between Nd-Co is found to be 3.11 \AA. Therefore, the hybridization between these two orbitals is very less expected. However, O $2p$ have broadened states and bond length of Co $3d$-O $2p$ (Nd $4f$-O $2p$) is 1.9 (3.1) \AA, which indicates the reasonable hybridization between them.
The resultant states from the hybridization of Nd-O and Co-O can be seen as indirect hybridization of Nd $4f$-Co $3d$ states.
This indirect hybridization can be the reason for the presence of finite Co $3d$ states around E$_f$. 
Further on adding the U$_f$$_f$, a creation of perfect band gap can be seen at $E_f$. 
In table 2, the columns \Romannum{2} and \Romannum{3}  are representing the band gap for spin-up and spin-dn states of the compound and column \Romannum{4} is representing the gap between two peaks of up-channel for Nd $4f$ in the vicinity of E$_f$ corresponding to varying U$_f$$_f$ value.
In column \Romannum{2} (\Romannum{3}), the band gap value of the compound at E$_f$ changed from 0 to 0.9 (1.27 to 1.02) eV for spin-up (dn) channel with U$_f$$_f$ varying from 0 to 5.5 eV.
In column \Romannum{4}, the gap increases from 0.24 to 4.8 eV with U$_f$$_f$ varying from 0 to 5.5 eV, which shows that Nd $4f$ states have shifted away from the E$_f$. 
The transport properties of a material are mainly decided by states around E$_f$ and thus for NCO the transport properties are supposed to be dominated by Co $3d$ states. 

\begin{table}\label{tabb}
\caption{\label{tabb}
\small{Varying band gap with U$_f$$_f$ of 0, 2.5, 3.5, 4.5, 5.5 eV, respectively alongwith U$_d$$_d$ of 2.7 eV.}} 
\begin{ruledtabular}
\begin{tabular}{lccc}
\textrm{U-Value/band gap} &
\textrm{Spin-up}& 
\textrm{Spin-dn}&
\textrm{Nd-4f-gap}\\ 
      
\colrule
$0$          & 0.0 & 1.27& 0.24 \\
$2.5$        & 0.68 & 1.04 & 2.33 \\
$3.5$        & 0.84 & 1.03 & 3.3\\
$4.5$        & 0.88 & 1.03& 4.43 \\
$5.5$        & 0.90 & 1.02 &4.8 \\

\end{tabular}
\end{ruledtabular}
\end{table}

\begin{figure} 
\begin{center}
\includegraphics[width=0.45\textwidth,height=6.0cm]{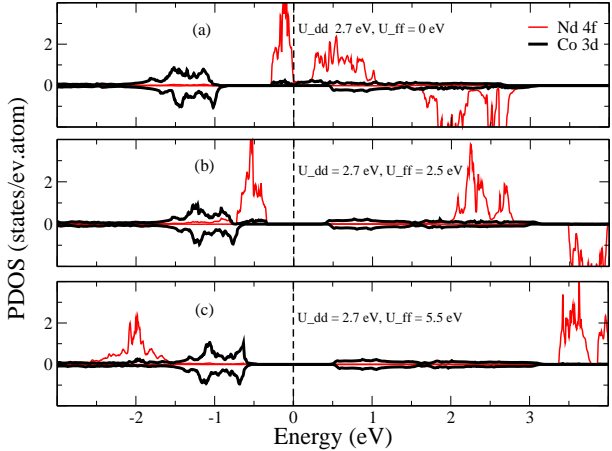}
\caption{\small{PDOS using DFT+$U$ calculation for U$_d$$_d$ = 2.7 eV and U$_f$$_f$ = 0, 2.5, 5.5 eV, respectively.
}}
\end{center}
\end{figure}


In order to analyse the effect of correlation, we have calculated the electronic dispersion curve for set of U$_f$$_f$ (U$_d$$_d$) values. 
Figure 6 a(b) is depicting the electronic dispersion of topmost valence band (VB) and bottommost conduction band (CB) for varying U$_f$$_f$ alongwith U$_d$$_d$ of 2.7 eV for spin-up (dn) states of the NCO. 
In figure 6(a), with increase in U$_f$$_f$ value from 2.5 to 5.5 eV, band gap of spin-up states increased from 0.68 to 0.90 eV.
 Also, band curvature at high symmetric k-points for U$_f$$_f$ = 3.5 to 5.5 eV found to be similar with each other, while it changed significantly for U$_f$$_f$ = 2.5 eV.
In figure 6(b), with increase in U$_f$$_f$ the band gap varies from $\sim$ 1.04 to 1.02 eV, that is band gap remains almost constant in spin-dn channel for the compound. 
Furthermore, for different U$_f$$_f$, the band curvature at high symmetric k-points for spin-dn states is found to be similar to each other.
One can infer from figure 6 a(b), that spin-up states have 0.36 eV less band gap in comparison to spin-dn states for U$_f$$_f$ = 2.5 eV. However, further increase in U$_f$$_f$ leads to decrement in the band gap difference between spin-up and spin-dn states to $\sim$ 0.1 eV.
This small difference in band gap for spin-up (dn) states, indicates that at higher temperature both spin would contribute in the transport behaviour of the compound. 
Since, the effective mass ($m^*$) plays important role in the magnitude of $\alpha$.  
Further, the maximum (minimum) energy of VB (CB) in figure 6 have been found  at high symmetric $\Gamma $-point, which indicates that compound have direct band gap for both spin-up and spin-dn channel.
Therefore, we have calculated the $m^*$ for both channels by using the energy band curvature around the $\Gamma $-point. 
The calculated values of $m^*$ for spin-up (dn) channel at $\Gamma $-point along high symmetric path of $\Gamma $-X, $\Gamma $-S, $\Gamma $-Z, $\Gamma $-R, and $\Gamma $-T are given in the table \Romannum{3}.
Here, the $\Gamma $-$X$ notation represents the calculated $m^*$ at high symmetric point $\Gamma $ along the $X$ direction.
 In table \Romannum{3} for spin-up states, the value of $m^*$ varies drastically in the VB when U$_f$$_f$ is increased from 2.5 eV to 3.5 eV and remains almost similar on further increase in U$_f$$_f$ value.   
 However, in CB no significant change in $m^*$ have been seen other than shifting of bands with varying U$_f$$_f$ value.
 Whereas, for spin-dn states no significant change in $m^*$  is seen for both VB and CB with varying U$_f$$_f$. 
From the table, one can see that the CB have dominating $m^*$ at $\Gamma $ along the $\Gamma $-Z direction with 30.99, 36.75, 36.75 (26.36, 25.96, 28.13) $m_e$ for spin-up (dn) states for varying U$_f$$_f$ = 3.5 to 5.5 eV.
It represents the electron as dominating charge carrier, indicates for the negative $\alpha $ value of the NCO. But, the experimentally measured value of $\alpha $ is positive, which is in contrast to above indicated negative $\alpha $.

\begin{figure*}
\includegraphics[width=0.9\linewidth, height=5cm]{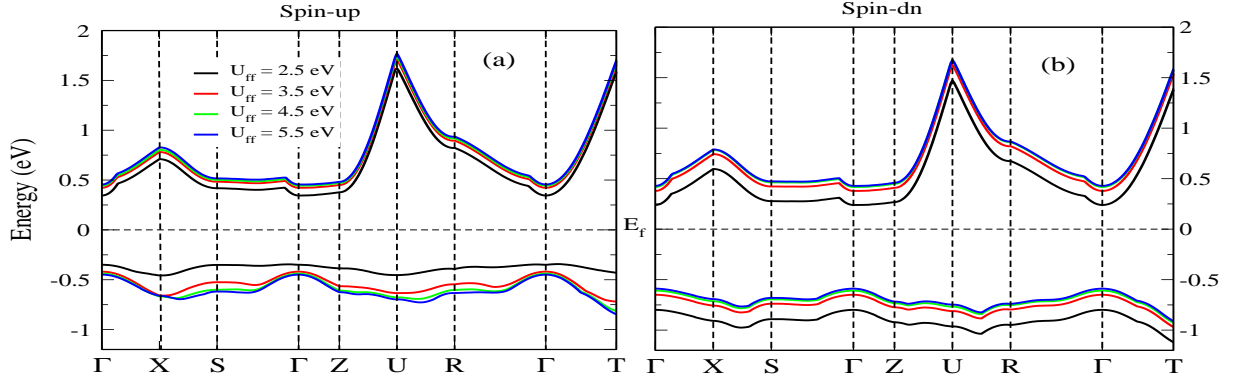}
\caption{\small{ Electronic dispersion curve of uppermost VB and lowermost CB of U$_d$$_d$ = 2.7 eV with varying U$_f$$_f$ = 2.5 to 5.5 eV for (a) Spin-up and (b) Spin-dn.}}
\end{figure*}

\begin{table*}\label{tabb1}
\caption{\label{tabb}
\small{The effective mass of holes and electrons at high-symmetric points.}} 
\begin{ruledtabular}
\begin{tabular}{lcccc}
 && \textrm{ Effective mass($m^*/m_e)$ }\\
\colrule
&&\textrm{{Spin-up  --- VB(CB)}}\\   
\colrule
\textrm{{K-point/U$_f$$_f$}}
&\textrm{{2.5 eV}}
&\textrm{{3.5 eV}}
&\textrm{{4.5 eV}}
&\textrm{{5.5 eV}}\\

\colrule
$\Gamma $-X    &-5.67 (0.83) & -3.66 (0.84)&  -3.57 (0.84) & -3.66 (0.82) \\
$\Gamma $-S    & 35.97 (0.93) & -3.36 (0.94)& -3.19 (0.94) & -3.19 (0.93) \\
$\Gamma $-Z    & -10.20 (21.28) &-3.26  (30.99) & -2.83 (36.75) & -2.68 (36.75)\\
$\Gamma $-R    & 17.02   (1.20) & -3.28 (1.22) &-3.02 (1.21) & -3.00 (1.19) \\
$\Gamma $-T     & 8.5 (1.34)  &-2.90 (1.38)& -2.65 (1.39) & -2.63 (1.36)\\
\colrule
&&\textrm{{Spin-dn --- VB(CB)}}
\\
        
\colrule
$\Gamma $-X    & -6.62 (0.89) & - 6.84 (0.87)& -6.92 (0.87) & -6.65 (0.89) \\
$\Gamma $-S    &  -5.38 (1.18) & -5.49 (1.03)&  -5.44 (1.03) & -5.29 (1.05) \\
$\Gamma $-Z    &  -3.39 (30.37) &-3.33 (26.36) & -3.31 (25.96) & -3.26 (28.13)\\
$\Gamma $-R    &   -4.67 (1.32) & -4.73 (1.29) &-4.68 (1.28) & -4.50 (1.31) \\
$\Gamma $-T     &-3.75 (1.59)  & -3.76 (1.54)&  -3.73 (1.53) &  -3.63 (1.63)\\

\end{tabular}
\end{ruledtabular}
\end{table*}


Density matrix is used to calculate the number of electrons for Nd $4f$, Co $3d$ and O $2p$ orbitals. Within the muffin tin radius, these orbitals have number of electrons equals 3, 6.4, and 3.4, respectively.
In pure ionic model, +3 (-2) state of Co (O) is expected and in this oxidation (reduction) state Co $3d$ (O $2p$) orbitals should have occupancy of 6 (6) electrons.
For O $2p$ orbitals, the obtained number of electrons is much less than from the expected. It is because oxygen have extended $p$-orbitals which can be beyond the muffin tin radius and PDOS is calculated inside the muffin tin radius only.
Moreover, 6.4 (3.4) occupancy for Co $3d$ (O $2p$) orbitals  and unoccupied states of O $2p$ orbitals are also indicating towards the $\textless $+3 ($\textless $-2) oxidation (reduction) states of the Co (O). 
This suggests the presence of covalent bonding between the Co and O atoms forming the octahedra. Which suggests the failure of pure ionic model and importance of covalent bonding in the compound.

\begin{figure*}
\includegraphics[width=0.85\linewidth, height=8cm]{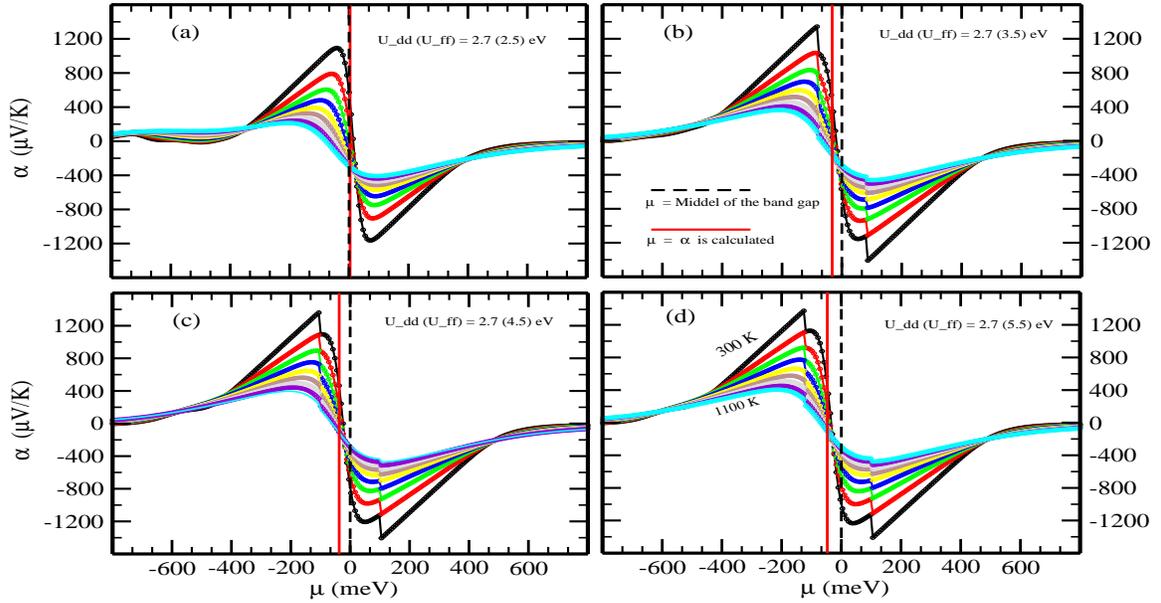} 
\caption{\small{Change in Seebeck coefficient ($\alpha $) with chemical potential ($\mu $) at different temperatures (using DFT + U for U$_f$$_f$ values of (a) 2.5 eV, (b) 3.5 eV (c) 4.5 eV (d) 5.5 eV alongwith U$_d$$_d$ = 2.7 eV).}}
\end{figure*}

In order to understand the experimentally measured $\alpha $ of the compound, we have calculated the $\alpha $ for spin-up (dn) channel by using the BoltzTraP\cite{boltz}. 
In the framework of two-current model, $\alpha $ has been calculated by using the formula 

\begin{equation}
\alpha  =\frac{\alpha_\uparrow \sigma_\uparrow+\alpha_\downarrow \sigma_\downarrow}{\sigma_\uparrow+ \sigma_\downarrow }
\end{equation}
where, $\sigma_\uparrow,_\downarrow$ ($\alpha_\uparrow,_\downarrow$) represents the electrical conductivity (Seebeck coefficient) for spin-up and spin-dn channels, respectively.
Figure 7 is showing the $\alpha$ versus $\mu $ in the temperature range of 300-1100 K.
 The U$_f$$_f$ used for figure 7 (a, b, c, and d) are 2.5, 3.5, 4.5, and 5.5 eV in combination with U$_d$$_d$ of 2.7 eV for each four cases. 
 In the figure dotted black line at the $\mu $ = 0 meV represents the middel of the band gap. The red line is representing the chemical potential ($\mu $) for which the calculated $\alpha$ best matches with the experimentally measured $\alpha$ at room temperature. 
One can see from figure that with increase in U$_f$$_f$, the rate of change of $\alpha$ becomes very sensitive with respect to $\mu $ at the vicinity of $E_f$. 
It is because with increase in U$_f$$_f$, the band gap increases and carrier concentration decreases, resulting in a substantial change in $\alpha $ value. 
At $\mu $ = 0 meV, the calculated values of $\alpha $ (at 300 K) are found to be 563, -610, -777, and -996 $\mu $V/K for U$_f$$_f$ of 2.5, 3.5, 4.5, and 5.5 eV, respectively.
Moreover, the obtained negative values of $\alpha $ for U$_f$$_f$ of 3.5, 4.5, and 5.5 eV are found to be consistent with the n-type behaviour as indicated by the estimated m$^*$ in the compound.
However, in our experimental study the value of $\alpha$ obtained at 300 K is positive, which shows the p-type behaviour in the NCO.
Here, one should keep in mind that $\mu $ changes with temperature in the compound, which suggests that the calculation of $\alpha $ at $\mu $=0 meV is not appropriate. 
The temperature dependence of $\mu $ in semiconducting materials is given by the following equation,\cite{ashcroft1}

\begin{equation}
\mu =\varepsilon_v + \frac{E_g}{2} +\frac{3k_BTln(m_v/m_c)}{4}
\end{equation}

where, $m_v$ and $m_c$ represents the effective mass (m$^*$) of the charge carrier in the VB and CB, $\varepsilon_v $ is the maximum energy of electron at VB, and E$_g$ represents the band gap.
Here, one should remember that $\varepsilon_v + \frac{E_g}{2}$ is middle of the band gap, i.e. $\mu $ = 0 meV.
 From table 3, using the maximum m$^*$ (for U$_f$$_f$ =5.5 eV and U$_d$$_d$ =2.7 eV) with value of  36.75 (-3.66) m$_e$ of electron (hole) for spin-up states, we get the value of $\mu $ = $\sim$ -44.7 meV at 300 K. 
 Similarly, by taking 28.13 (-6.65) m$_e$ as m$^*$ value for electron (hole) in spin-dn states, the value of $\mu $ found to be -27.98 meV at room temperature.
 Therefore, it is important to notice that value of $\mu $ at room temperature will be less than -27.98 meV at thermal equilibrium.
The calculated values of $\alpha $ in figure 7(d) at $\mu $ less than -27.98 meV shows the hole dominating behaviour in the compound, 
 which is in agreement with our and Tealdi $et$ $al.$ work.

Figure 8 shows the $\alpha$ with varying temperature.  
In order to get the best match of theoretical $\alpha $ with the expriment at room temperature, for each four cases of figure 7 (a-d), the $\mu $ is shifted by 3.20, -32.44, -37.13, -46.98 meV, respectively from the $\mu $ = 0 meV to get the experimentally measured $\alpha $ value of 444 $\mu $V/K at 300 K.
Furthermore, using these $\mu $ values, we have calculated the $\alpha $ in the temperature range of 300-600 K, which is shown in the figure 8.   
 Figure shows that the calculated $\alpha $ value has the lowest difference for U$_f$$_f$ of 5.5 eV when compared to the experimental data.
 In addition, the difference in $\alpha $ value between best match of calculated and experimental values determined to be $\sim$ 95, 180, 203, 186, 149, 105 $\mu $V/K at temperatures of 350, 400, 450, 500, 550, 600 K, respectively.
The maximum difference between calculated and experimentally obtained $\alpha $ value is $\sim$ 203 $\mu $V/K at 450 K, which decreases with the rise in temperature.
Here, the observed difference between experimental and calculated $\alpha $ may be due to various factors, such as spin state transition, relaxation time, and temperature dependent $\mu $\cite{PhysRevB.61.5303,PhysRevB.81.165203,Jambunathan}. 
Above, we have discussed the $\alpha $ vs $T$ for fixed $\mu $ at room temperature. 
Further, for U$_f$$_f$ = 5.5 eV case, we simply select those $\mu $ at each temperature for which calculated $\alpha$ can always get the matching value with experimental $\alpha$. 
 The different $\mu $ chosen at temperature 300, 400, 500, and 600 K  are  $\sim$ -46.98, -55.26, -58.50, -55.21 meV, which corresponds to p-type doping of 0, 5.0$\times$$10^1$$^4 $, 9.6$\times$$10^1$$^5 $, 5.2$\times$$10^1$$^6 $ cm$^-$$^3$, respectively. 
   
\begin{figure}
\includegraphics[width=0.98\linewidth, height=7cm]{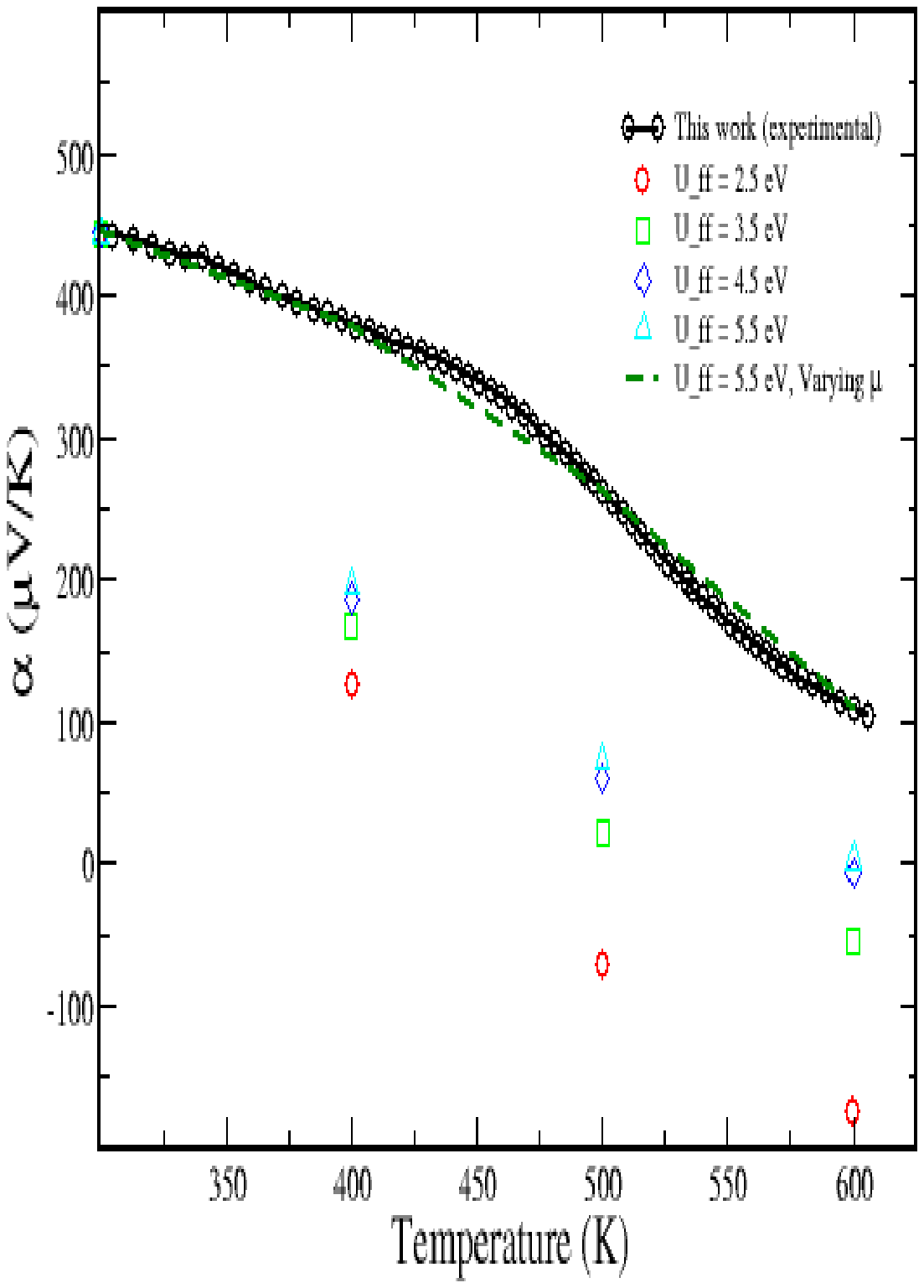} 
\caption{\small{Comparison of experimental and calculated (using DFT + U for $U_d$$_d$ = 2.7 eV with varying $U_f$$_f$ from 2.5 - 5.5 eV) values of Seebeck coefficient ($\alpha $) as a
function of temperature.}}
\label{}
\end{figure}

\begin{figure}
\includegraphics[width=0.95\linewidth, height=6.5cm]{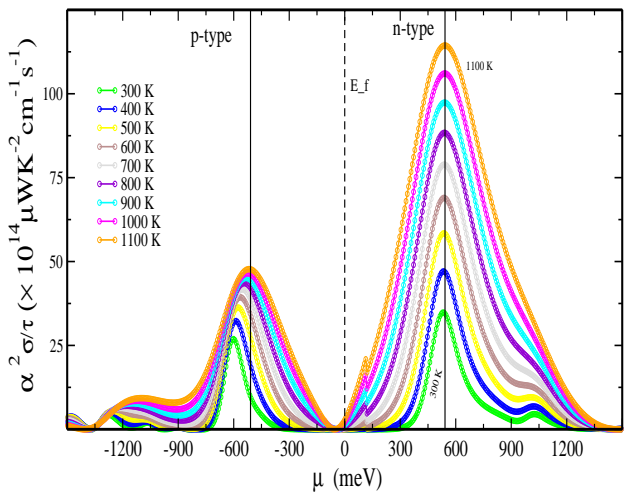} 
\caption{\small{Variation of Power Factor (PF) with chemical potential ($\mu $) at different temperatures.}}
\label{}
\end{figure}

Enhancing the power factor (PF) by doping in the TE materials is one of strategies to achieve the higher figure of merit.
Therefore, theoretically predicting the doping of suitable carrier concentration through the PF versus $\mu $ plot would be helpful in achieving the higher PF.
In order to analyse the TE potential capabilities of the compound, we have shown the variation of PF versus $\mu $ with varying temperature from  300-1100 K in figure 9 for U$_f$$_f$ = 5.5 eV, U$_d$$_d$ = 2.7 eV.
Consecutive temperatures are to be read from 300-1100 K in order of their peak height.
Each curve is representing different temperature, PF value is maximum at 1100 K corresponding to two peaks for negative and positive values of $\mu $.
The negative and positive values of $\mu $ is representing the holes and electrons type doping, respectively. 
For the negative values of $\mu $, the maximum PF at 300 K found to be $\sim$ 26.8 $\times$ $10^1$$^4 $ $\mu $$W$$K^-$$^2$$cm^-$$^1$$s^-$$^1$ at $\sim$ -600.2 meV.
 With increase in temperature from 300 K to 1100 K, we observe a shift in $\mu $ from $\sim$ -600.2 to -509.1 meV to achieve the maximum PF of $\sim$ 47.6 $\times$ $10^1$$^4 $ $\mu $$W$$K^-$$^2$$cm^-$$^1$$s^-$$^1$ which corresponds the hole doping of $\sim$ 1.4 $\times$ 10$^2$$^1$ cm$^-$$^3$.
For positive values of $\mu $, the maximum observed PF is $\sim$ 34.8 $\times$ $10^1$$^4 $ $\mu $$W$$K^-$$^2$$cm^-$$^1$$s^-$$^1$ at $\sim$ 535 meV at 300 K. In addition, at 1100 K, the PF found to be $\sim$ 114.4 $\times$ $10^1$$^4 $ $\mu $$W$$K^-$$^2$$cm^-$$^1$$s^-$$^1$ at $\sim$ 542 meV with very small shift in $\mu $ of $\sim$ 7 meV from 300-1100 K. In order to achieve the maximum obtainable PF at $\mu $ = $\sim$ 542 meV, the required electron doping is $\sim$ 0.7 $\times$ 10$^2$$^1$ cm$^-$$^3$. 
Therefore, to achieve the maximum power factor, n-type doping will be required for the compound to enhance its thermoelectric behaviour.

\section{Conclusions}
In conclusion, we have analysed the $\alpha $ of NCO using experimental and DFT+$U$ approach in the high-temperature region.  
The measured values of $\alpha $ obtained at 300 (600) K is $\sim $444 (109.8) $\mu $V/K. 
Furthermore, with increase in temperature from 300-600 K, continuous decrease in $\alpha $ is observed. In addition, the rate of decrement of the $\alpha $ below 460 K found to be $\sim$ -0.73 $\mu V/K^2$, which increases to $\sim$ -1.56 $\mu V/K^2$ in the temperature range of 460-600 K.
The occupied states 6.4 of Co $3d$ and large unoccupied states of O $2p$ indicates the presence of covalent bonding in the compound.
In addition, the m$^*$ around $\Gamma $ along the $\Gamma $-Z direction in CB found as maximum which indicates the electron as dominating charge carrier in NCO. However, calculated value of $\alpha $ shows the p-type behaviour of the NCO after considering temperature dependence in $\mu $. 
  SP DFT+$U$ (U$_f$$_f$ = 5.5 eV, U$_d$$_d$ = 2.7 eV) calculation gives the best match of $\alpha $ value when compared to the experiment.
  Maximum PF is calculated for both p and n-type behaviour of the compound.  For p (n)-type, the maximum PF is found to be 47.6 (114.4) $\times$  $10^1$$^4 $ $\mu $$W$$K^-$$^2$$cm^-$$^1$$s^-$$^1$ at 1100 K, which corresponds to p (n)-type doping of $\sim$ 1.4 (0.7) $\times$ 10$^2$$^1$ cm$^-$$^3$. 
  In summary, we have systematically investigated the key role of the strong electronic correlation of Nd $4f$ and Co $3d$ orbital and its role in the $\alpha $. 
The present study shows the efficient path for investigating the thermoelectric properties in rare earth cobaltates. 
\bibliography{ref}
\bibliographystyle{apsrev4-1}

\end{document}